\documentclass[twocolumn]{webofc}
\usepackage[varg]{txfonts}   
\usepackage{graphicx}
\graphicspath{%
  {./},%
    {./Archiv/},%
    {Archiv/},%
}
\begin{document}
\title{Towards a first measurement of the free neutron bound beta decay detecting hydrogen atoms at a throughgoing beamtube in a high flux reactor}
%
%

\author{\firstname{Wolfgang} \lastname{Schott}\inst{1}\fnsep\thanks{\email{wolfgang.schott@tum.de}} \and
        \firstname{Erwin} \lastname{Gutsmiedl}\inst{1}
	\and
        \firstname{Karina} \lastname{Bernert}\inst{1}
	\and
        \firstname{Ralf} \lastname{Engels}\inst{2}
        \and
        \firstname{Roman} \lastname{Gernh\"auser}\inst{1}
	\and
        \firstname{Stefan} \lastname{Huber}\inst{1}
        \and
        \firstname{Igor} \lastname{Konorov}\inst{1}
	\and
        \firstname{Bastian} \lastname{M\"arkisch}\inst{1}
	\and
        \firstname{Stephan} \lastname{Paul}\inst{1}
        \and
        \firstname{Christoph} \lastname{Roick}\inst{1}
	\and
        \firstname{Heiko} \lastname{Saul}\inst{1}
	\and
        \firstname{Suzana} \lastname{Spasova}\inst{1}
}

\institute{Physik-Department, Technische Universit\"at M\"unchen, D-85748 Garching, Germany
\and
           Institut f\"ur Kernphysik, Forschungszentrum J\"ulich, D-52425 J\"ulich, Germany
          }

\abstract{In addition to the common 3-body decay of the neutron $n\rightarrow p e^-\overline{\nu_e}$ there should exist an effective 2-body subset with the electron and proton forming a Hydrogen bound state with well defined total momentum, total spin and magnetic quantum numbers. The atomic spectroscopic analysis of this bound system can reveal details about the underlying weak interaction as it mirrors the helicity distributions of all outgoing particles. Thus, it is unique in the information it carries, and an experiment unravelling this information is an analogue to the Goldhaber experiment performed more than 60 years ago. The proposed experiment will search for monoenergetic  metastable BoB H atoms with 326 eV kinetic energy, which are generated at the center of a throughgoing beamtube of a high-flux reactor (e.g., at the PIK reactor,
Gatchina). Although full spectroscopic information is needed to possibly reveal new physics our first aim is to prove the occurrence of this decay and learn about backgrounds. Key to the detection is the identification of a monoerergtic line of hydrogen atoms occurring at a rate of about 1 $\rm{s}^{-1}$ in the environment of many hydrogen atoms, however having a thermal distribution of about room temperature. Two scenarios for velocity (energy) filtering are discussed in this paper. The first builds on an purely electric chopper system, in which metastable hydrogen atoms are quenched to their ground state and thus remain mostly undetectable. This chopper system employs fast switchable Bradbury Nielsen gates. The second method exploits a strongly energy dependent charge exchange process of metastable hydrogen picking up an electron while traversing an argon filled gas cell, turning it into manipulable charged hydrogen. The final detection of hydrogen occurs through multichannel plate (MCP) detector.
The paper describes the various methods and gives an outlook on rates and feasibility at the PIK reactor in
Gatchina.}
\maketitle
\setlength{\parindent}{0em}
\section{Introduction}
\begin{figure}[ht]
\includegraphics[width=0.49\textwidth]{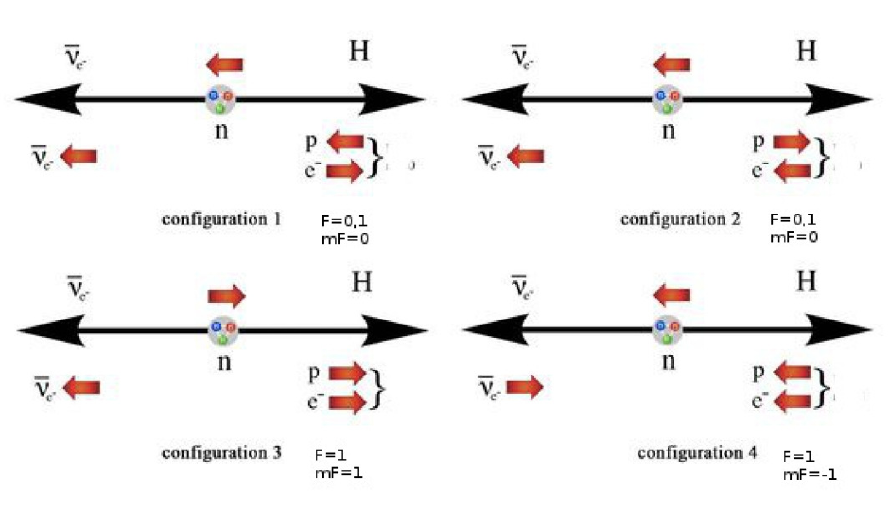}
\caption{Neutron bound beta decay hydrogen atom hyperfine states (black: momentum, red: spin). Configurations 1, 2, 3 are allowed within the V-A theory.   For the configuration 4 a right-handed neutrino is needed. $F$ is the total hyperfine spin and $m_F$ the $F$ projection.}
\label{fig.1}
\end{figure}

The neutron decay has for many years been and is subject of intense studies, as it reveals detailed information on the structure of the weak interaction \cite{Dubbers1}. Using the two-body neutron decay into a hydrogen atom and an electron anti-neutrino
$n \rightarrow$ \rm{H} $+ \bar{\nu_e}$, the hyperfine populations of the emerging hydrogen atom can be investigated \cite{Nem1}. The challenge lies in the very small $4 \cdot 10^{-6}$ branching ratio to the total neutron decay rate. Hydrogen atoms from this decay have 325.7 eV kinetic energy corresponding to a $\beta=v/c$ of $0.83 \cdot 10^{-3}$ (non-relativistic Hydrogen atom).
Due to conservation of angular momentum, the electron populates only s-states in the Hydrogen atom ($ 83.2\% $ H(1s), $10.4\%$ H(2s)).
If one applies the standard purely left handed V-A interaction (the antineutrino helicity $H_{\bar{\nu}}$ being 1) \cite{Nem2,Schott1}, three of the four possible hyperfine spin states are allowed (see Fig. \ref{fig.1}).

\
Their populations are given by

\begin{equation}
w_1=\frac{(\chi -1)^2}{2(\chi^2+3)}
\label{eq.1}
\end{equation}

\begin{equation}
w_2=\frac{2}{\chi^2+3}
\label{eq.2}
\end{equation}

\begin{equation}
w_3=\frac{(\chi +1)^2}{2(\chi^2+3)}
\label{eq.3}
\end{equation}

\begin{equation}
\chi=\frac{1+g_S}{\lambda -2g_T}.
\label{eq.4}
\end{equation}

They depend on $\lambda =\frac{g_A} {g_V}$ (ratio of the weak axial-vector and vector coupling constants of the nucleon, $\lambda$ = -1.27641$\pm$ 0.00056 \cite{Maerkisch}), and also on
the scalar and tensor coupling constants $g_S$ and $g_T$.
Thus by measuring the populations $w_1$ - $w_3$ of these spin states, a combination of $g_S$ and $g_T$ can be obtained.
The population $w_4$ of the spin state shown as configuration 4 in Fig. \ref{fig.1} can only occur, if right-handed neutrinos are emitted \cite{Byrne}.
Applying the left-right symmetric model with its V+A admixture, leads to

\begin{equation}
w_4=\frac{(x+y\lambda)^2}{2\cdot (1+3\lambda ^3+x^2+3\lambda ^2y^2)},
\label{eq.5}
\end{equation}

where $x=\eta-\varsigma$, and $y=\eta+\varsigma$ \cite{Byrne}. The parameter $\eta$ depends on the mass ratio of two intermediate charged vector bosons, and $\varsigma$ is the mixing angle of the boson's mass eigenstates.
From the $\mu^+$ decay, upper limits for these parameters can be deduced \cite{Gap,Mus} ($\eta<0.036$ and $\varsigma<0.03$ ; C.L. 90$\%$).
The antineutrino helicity can be expressed by

\begin{equation}
H_{\bar{\nu}}=\frac{1+3\lambda^2 -x^2-3\lambda^2y^2}{1+3\lambda^2+x^2+3\lambda^2y^2}\leq1.
\label{eq.6}
\end{equation}

If $H_\nu<1$ is measured, i. e., right-handed neutrinos obviously occur in the experiment, a left-right  symmetric extension of the Standard Model may apply, where a new heavy gauge boson $W_2$ occurs, which couples to right- handed particles and the known $W_1$ boson primarily coupling to left-handed particles. $\eta$ is the $W_1$  over $W_2$ mass ratio squared. By linear combination of $W_1$ and $W_2$ with the mixing angel $\zeta$, left- and right-handed mass Eigenstates result \cite{Byrne}.

If one sets $\varsigma$=0 and $\eta$=0.036, then the population of the forbidden spin state becomes w$_4 \simeq$10$^{-5}$, and the helicity of the antineutrino $H_{\bar{\nu}}$=0.997.
The goal of the planned BoB experiments is to reduce the upper limit of $|g_S|$<6$\cdot$10$^{-2}$ (C.L. 68\%) \cite{Ade} by a factor of 10. The helicity of the antineutrino should be determined with an accuracy of 10$^{-3}$. With this accuracy one can set the statistical uncertainty of $\eta$ to $\simeq$10$^{-2}$, and therefore via Eq. \ref{eq.5} also the necessary statistical uncertainty of $w_4$.

\begin{figure}[ht]
\includegraphics[width=0.49\textwidth]{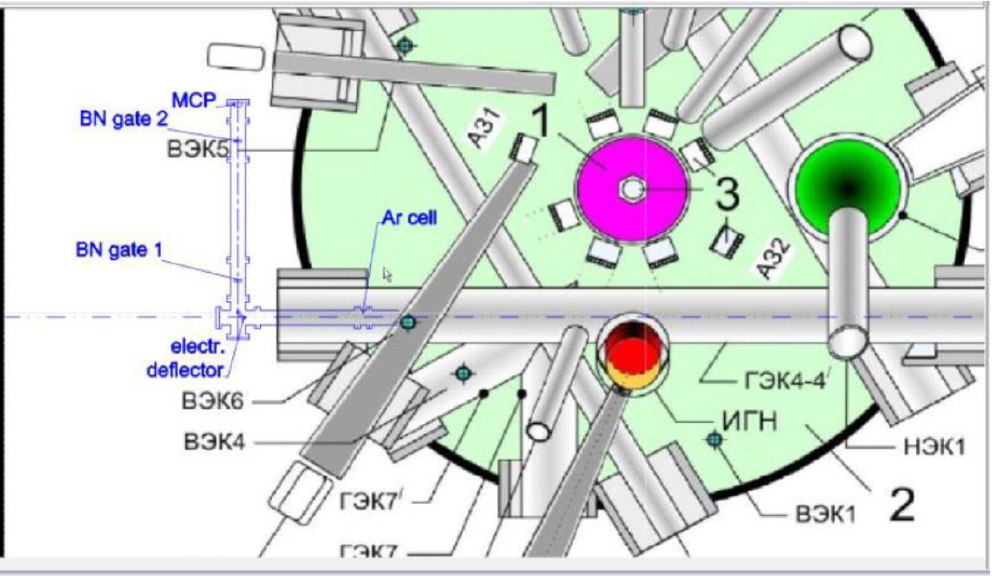}
\center
\caption{BoB setup at the Gatchina PIK reactor consisting of an Ar gas cell, an electrostatic focusing element, a pulsed electric deflector, a BN gate chopper and a MCP detector.}
\label{fig.2}
\end{figure}

It is planned, to perform the first experiments at the PIK reactor in Gatchina (Russia). In a first setup (see Fig. \ref{fig.2}) we will install in one of the throughgoing beam tubes of this reactor an Argon gas cell, located in the high neutron flux area, close to the fuel element. The metastable hydrogen atoms H(2s) will capture an electron from the Argon atoms, and be transformed to H$^-$ ions. These ions have almost the same energy as the H(2s) atoms ($\simeq$326 eV). The cross section for electron capture by H(2s) is roughly two orders of magnitude larger, compared to the reaction with hydrogen atoms in the ground state (H(1s)) \cite{Argon}. During the capture process, the 2s state decays, and the energy difference between the 2s and 1s state is transferred to the H$^-$ ion ($\simeq$10 eV \cite{Schott2}) as gain in kinetic energy.

Inside the throughgoing beam tube, a combination of Einzel lenses \cite{Einzel} will focus the H$^-$ ions  onto
the entrance of a pulsed electrical deflector, outside of the biological shield of the reactor. This deflector (see Fig. \ref{fig.3}) will bend the H$^-$ ions by $90^o$ out of the direct view into the throughgoing beam tube.
This strongly reduces the direct background coming from the beam tube.

The H$^-$ ions pass subsequently through two Bradbury Nielsen gates (BN gates), which work as an electrical time-of-flight (TOF) system \cite{BNG1} for charged particles, enabling us to measure the energy of the H$^-$ ions with a resolution of about 1.6$\%$.
The H$^-$ ions are counted by a multi-channel-plate (MCP) detector afterwards.
A second  method of determining the velocity/energy of the H$^-$ ions would be the counter-field method, which is described in detail in Ref. \cite{Schott2}.
Detailed simulation of the background, emerging from the throughgoing beam tube still have to be performed (MCNP \cite{MCNP}) for the PIK reactor. We have already done such simulations for the situation at the beam tube SR6 at the FRM II reactor in Munich. We developed for that beam tube a concept (shielding, collimation), which reduces the neutron and $\gamma$ background to a level at which the BoB experiment becomes feasible. These results will be published elsewhere. Furthermore, we have investigated the effects of residual gases in the beam tube. It turned out, that a cooled insert tube in the throughgoing beam tube is necessary, in order to freeze out the gas molecules, which otherwise would disturb the traveling H(2s) atoms.
\par
A second detection scenario for bound neutron $\beta$-decays is the use of a velocity filter for metastable hydrogen atoms, again using a system of two switchable gates in the beam tube. These gates act on the 2s state and will quench metastable hydrogen by means of an electric field in the closed mode and leave it untouched in the open mode. Placing two fast switching gates at fixed distance with appropriate phase shift of open/close state acts as a narrow band velocity filter (chopper). The detection of surviving metastable hydrogen atoms can proceed with an argon cell described above or by means of a quenching plate mounted close to a MCP to detect electrons released from the plate in this process.

\section{Ongoing Work}
\label{sec-1}

\subsection{Bradbury Nielsen gate chopper}
\label{sec-2}
\begin{figure}[h]
\includegraphics[width=0.49\textwidth]{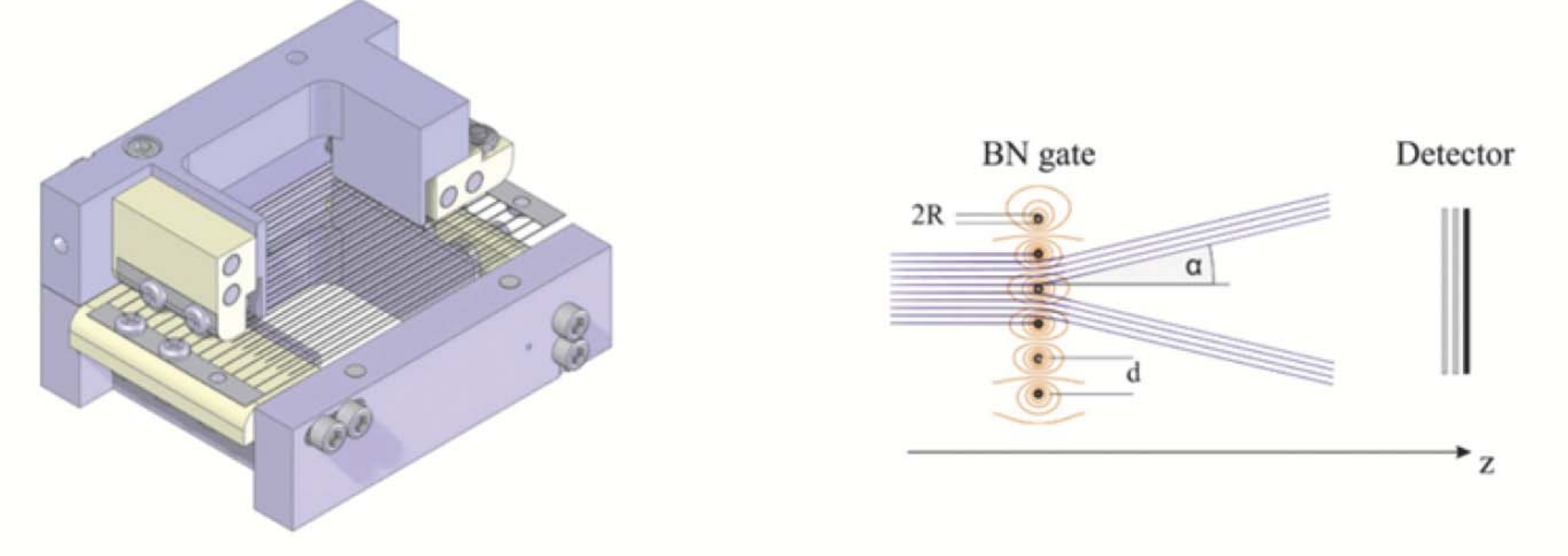}
\center
\caption{Left: A design drawing of a BN gate. Right: BN gate section view and functioning. The device consists of two insulated grids with equal + and - voltages, producing an electric field between the grid wires by means of which charged particles are deflected. At zero voltage, the particles pass the gates non- deflected. Neutral particles, e.g., metastable H(2s) hydrogen atoms, are de-excited into H(1s) by means of the electric field. At zero field they remain to be H(2s). }
\label{fig.4}
\end{figure}

\begin{figure}[h]
\includegraphics[width=0.49\textwidth]{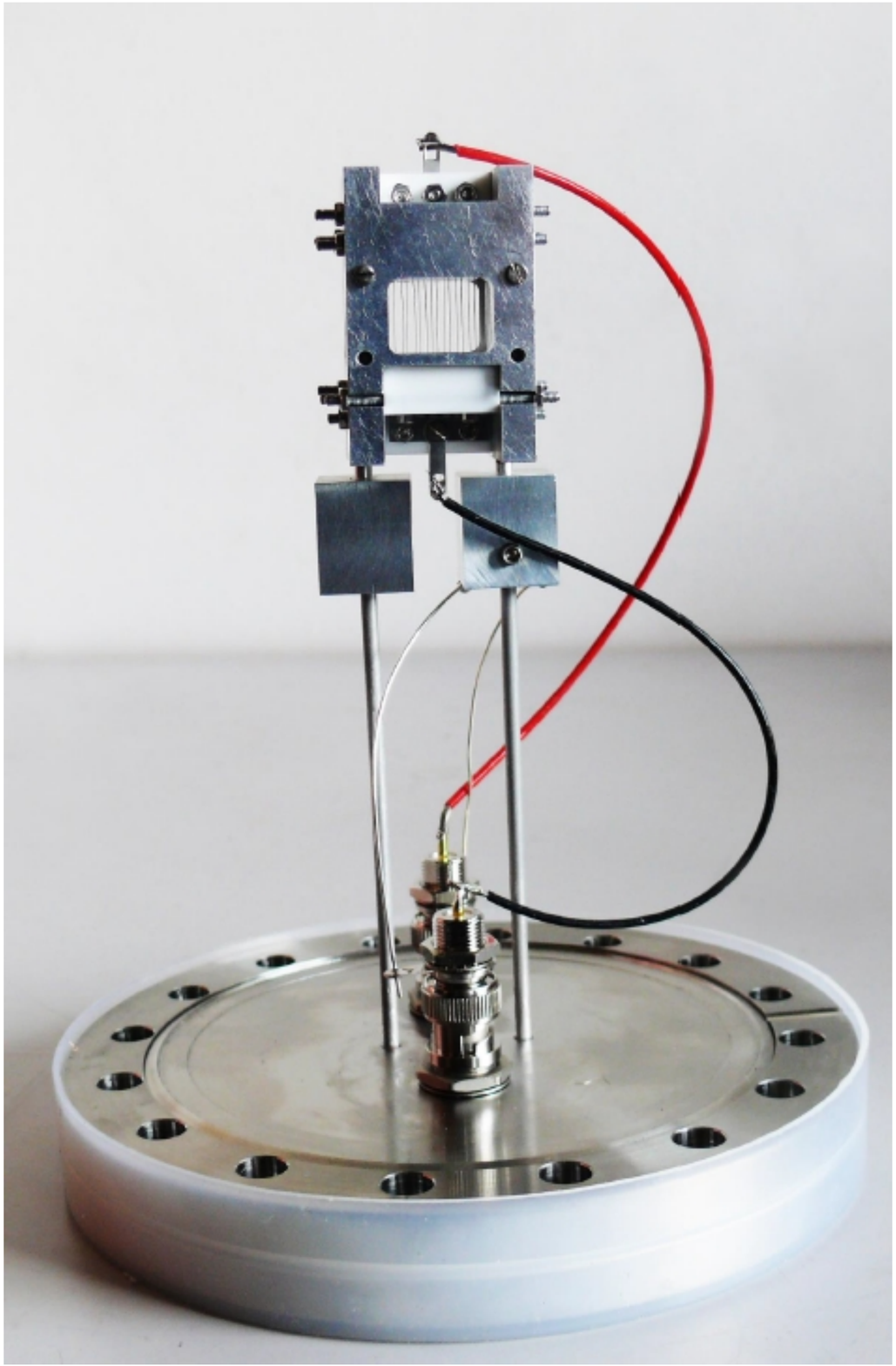}
\center
\caption{Bradbury Nielsen gate consisting of two insulated wire grids being chargeable with $\pm$ 500\ V each. For adjusting the grid input resistance to the 50$\ \Omega$ cable between electronics and BN gate, the grids are grounded by a serial RC element with $R=50\ \Omega$ and $C=100\ nF$ (not shown).}
\label{fig.5}
\end{figure}

\begin{figure}[h]
\includegraphics[width=0.49\textwidth]{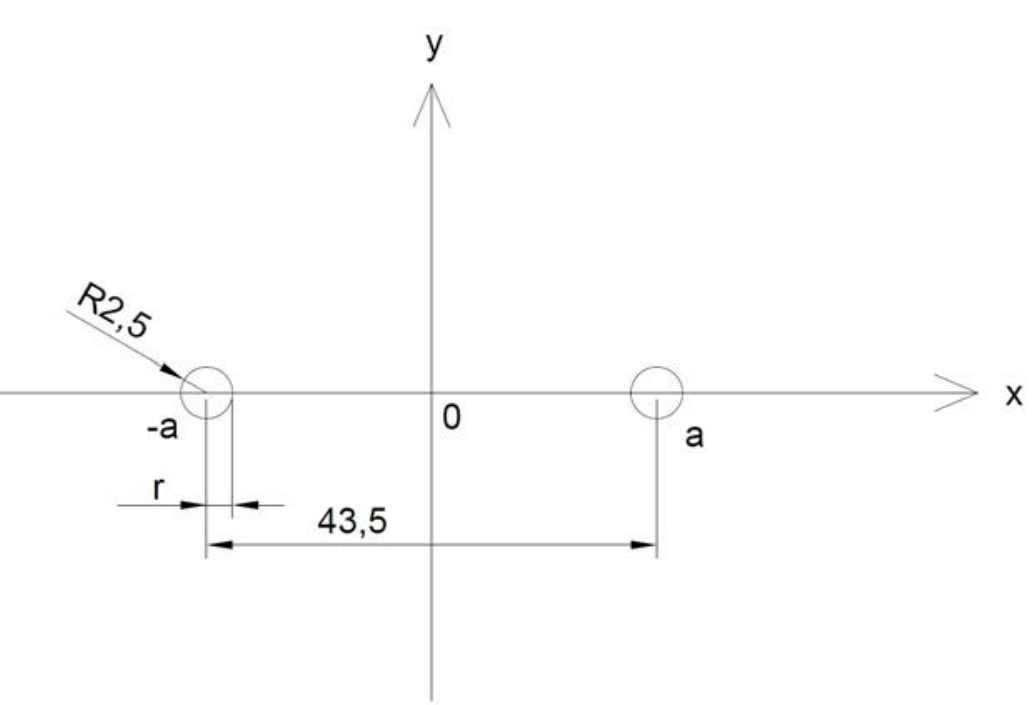}
\center
\caption{The positively and negatively charged wires of a BN gate belong to two concatenated grids. The schematic shows the geometry of adjacent wires, with dimensions in mil.}
\label{fig.6}
\end{figure}

\begin{figure}[h]
\includegraphics[width=0.49\textwidth]{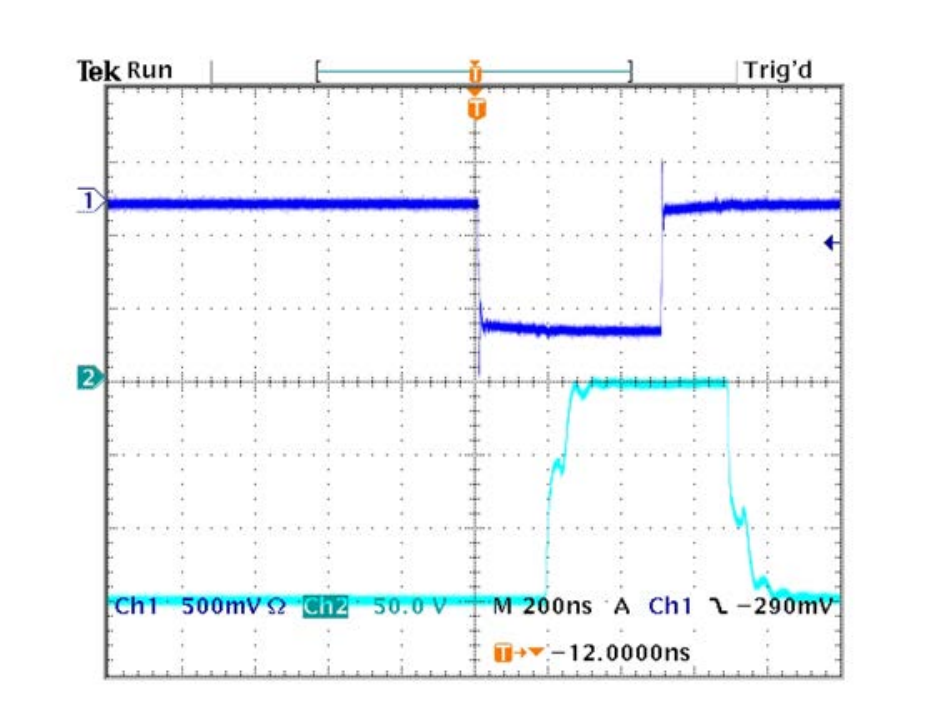}
\center
\caption{Signals measured using an oscilloscope. Blue: BN gate trigger NIM signal. Turquoise: -150\ V BN gate pulse.}
\label{fig.7}
\end{figure}

A BN gate consists of a layer of wires mounted on insulating frames as shown in Fig. \ref{fig.4}. Opposite electric potentials applied to adjacent wires generate local electric fields between them which deflect charged particles out of the beam, as shown at the right in Fig. \ref{fig.4}. Switching off the voltage removes the deviating effect. One can therefore set up a time-of-flight (TOF) system, using two BN gates at a certain distance, combined with a fast switching electronics \cite{BNG1}. While in the first step of the neutron BoB experiment this concept will be applied for H$^-$ ions, it works also for metastable H(2s) atoms which are quenched to the 1s state in the vicinity of a charged wire and thus are no longer available for study in the beam.
A typical pulse signal applied to a BN gate using the electronic system developed in-house is shown in Fig. \ref{fig.7}.
A typical pulse signal for one BN gate is shown in Fig. \ref{fig.7}. Switching BNG1 and BNG2 (see Fig. \ref{fig.10}) with short pulses (ns), and with a delay for the second gate, leads to a selection of a defined velocity of the charged particle or H(2s) atom with a good energy resolution of a few percent.
Figure \ref{fig.5} depicts one of our BN gates. Its aperture is 1.76 cm $\times$ 1.26 cm.
The geometrical dimensions of the BN grid wires are shown in Fig. \ref{fig.6}.

\begin{figure}[h]
\includegraphics[width=0.49\textwidth]{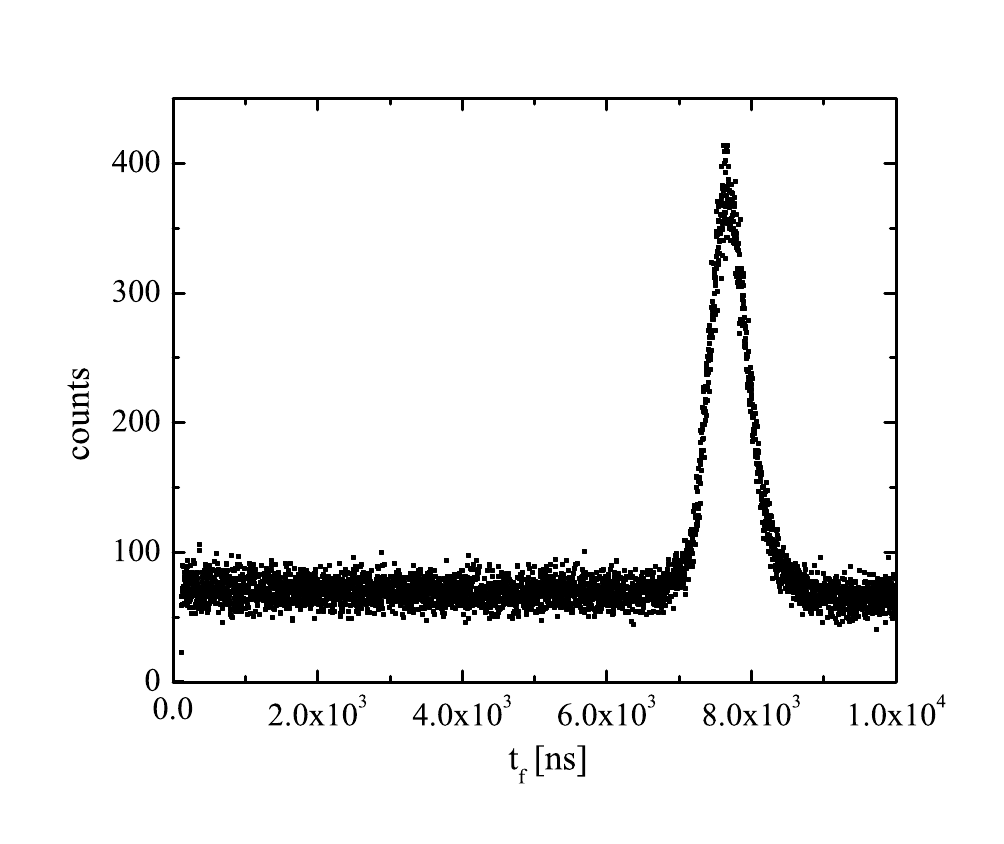}
\center
\caption{ TOF spectrum of 500 eV protons.  The delay between the pulses applied to the BN gates is 3.3 $\mu$s, while the delay $t_{f}$ between the pulse to BNG1 and the detected signal is 7.2 $\mu$s.}
\label{fig.8}
\end{figure}

For example, if a voltage of $U_{+/-}=\pm\ 200\ \rm{V}$ is applied between the wires, an electrical field of  $E=2.5\cdot\ 10^5\ \rm{V/m}$ is produced at $(x,y) = (0,0)$.
A proton with 500 eV energy will be deflected by this field with an angle of 14.7$^o$ out of the collimated beam direction.
As proof of principle we tested our BN gate system with 500 eV protons, coming from a strong plasma source (see next section for details).
The TOF spectrum is shown in Fig. \ref{fig.8}. The gates were switched with 500 ns pulses (duty time), which determine the FWHM of the peak.

\subsection{A proton/ion source for R$\&$D experiments towards BoB}
\label{sec-3}

\begin{figure}[h]
\includegraphics [scale=0.06]{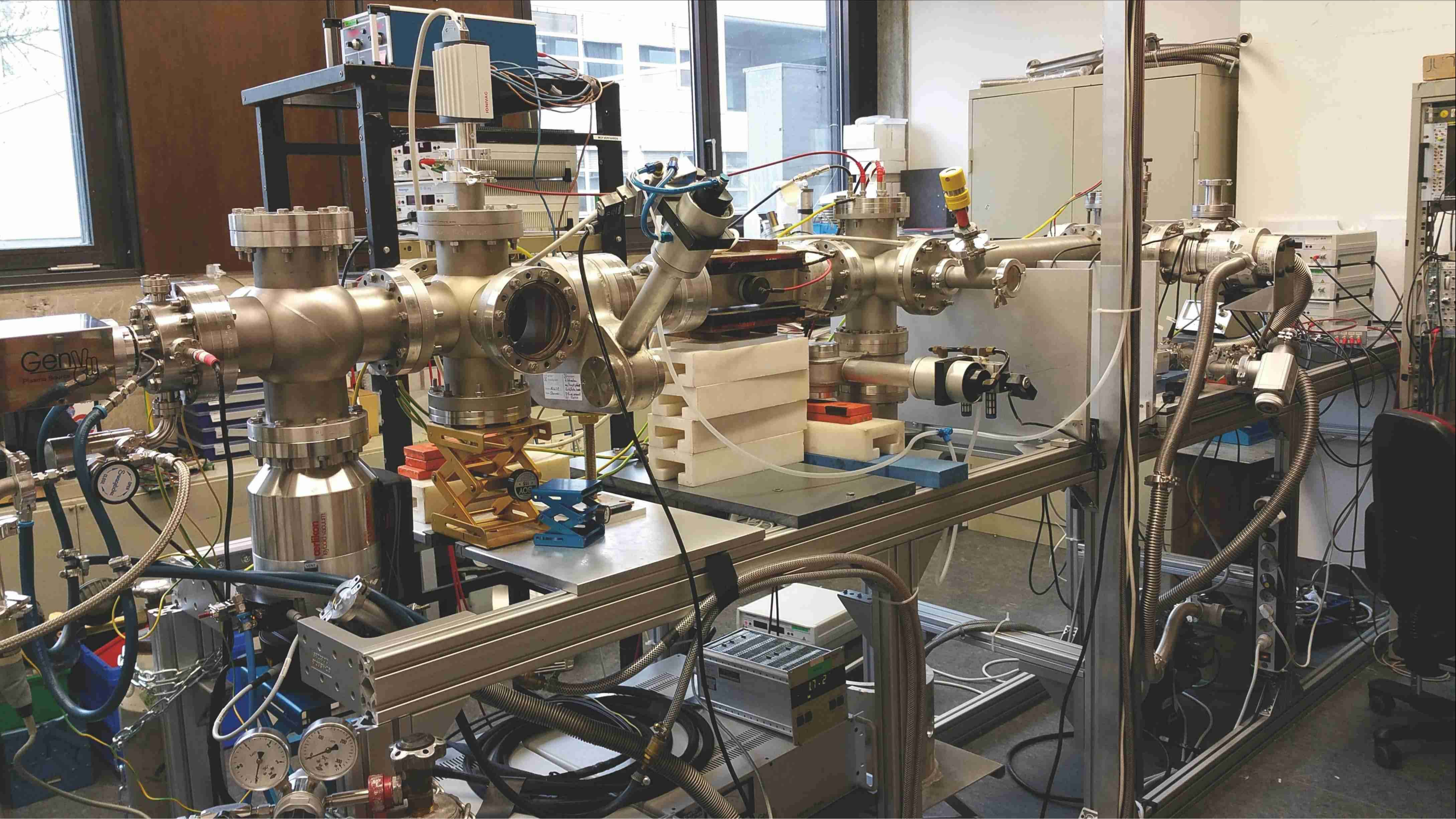}
\center
\caption{Test facility in BoB laboratory at TUM.}
\label{fig.9}
\end{figure}

\begin{figure}[h]
\includegraphics[width=0.49\textwidth]{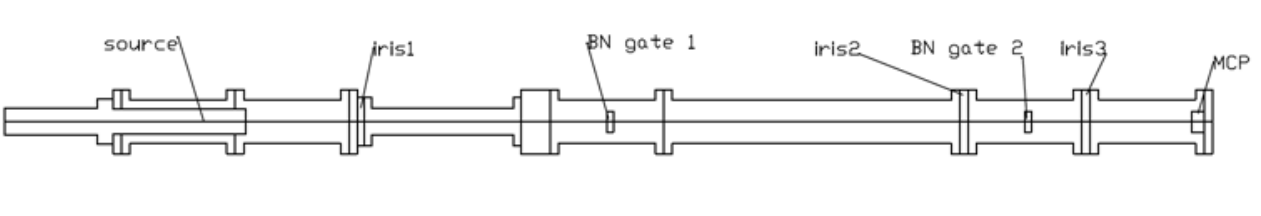}
\center
\caption{Sketch of an experimental setup at the ion source in the BoB lab.}
\label{fig.10}
\end{figure}

For the purpose of developing the necessary tools for a future BoB experiment, we have built an experimental test facility with a
commercial ion source, which is normally used in the semiconductor production \cite{TECTRA} (see Fig. \ref{fig.9}).
We have set up a beam line for protons and ions, which are collimated by iris diaphragms. Figure \ref{fig.10} shows a typical setup of our experiments.

\begin{figure}[h]
\includegraphics[width=0.49\textwidth]{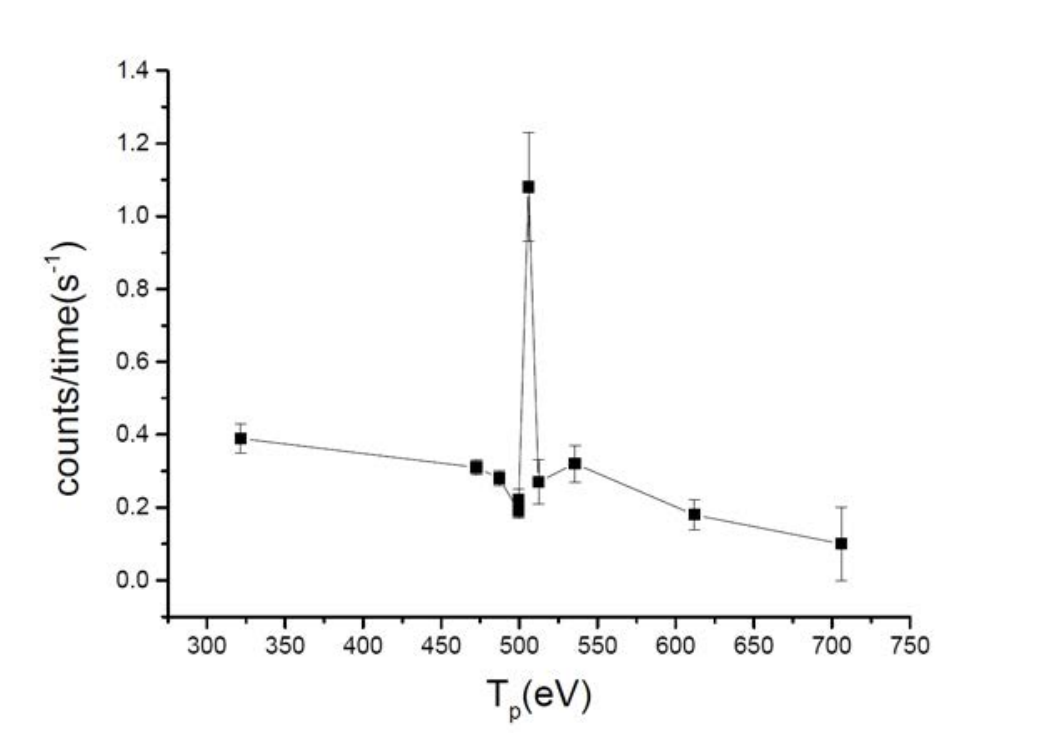}
\center
\caption{Plasma proton source line profile at $5\cdot 10^{-4}$\ mbar source $\rm{H}_2$ pressure, measured by varying the delay time between the gates and, thus, scanning the spike $T_p$ over the line width. The error bars denote the statistical error.}
\label{fig.11}
\end{figure}

\begin{figure}[h]
\includegraphics[width=0.49\textwidth]{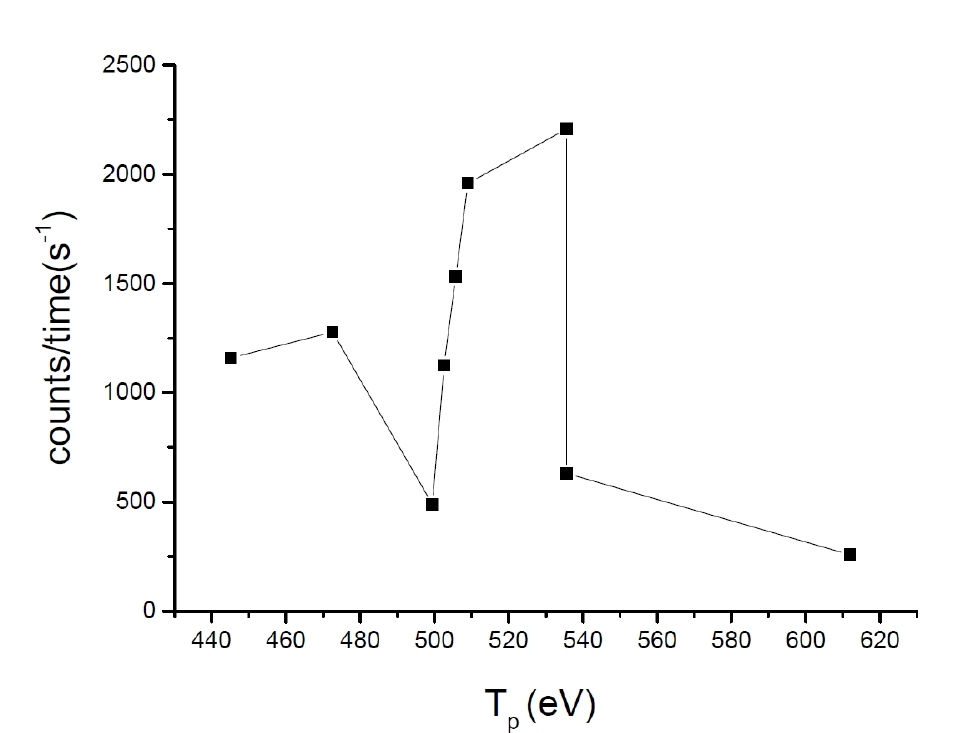}
\center
\caption{Proton source line profile at $5\cdot 10^{-3}$\ mbar source $\rm{H}_2$ pressure.}
\label{fig.12}
\end{figure}

If the plasma source is driven with hydrogen gas, a stable proton beam can be produced. The energy of this beam can be set by an extraction voltage. If the pressure in the plasma source is not too high an almost mono-energetic beam can be achieved. Figure \ref{fig.11} and \ref{fig.12} show the plasma source proton line profile for two different pressures.
The proton source peak at low pressure ($5\cdot 10^{-4}$\ mbar) has roughly a FWHM of 10 eV. The manufacturer of the source (TECTRA) quotes an intrinsic FWHM line width of 10 eV at a pressure of $5\cdot\ 10^{-4}$\ mbar. At higher pressure the FWHM line width of the proton source line profile increases, e.g., to $\sim$ 30 eV at $5\cdot 10^{-3}$\ mbar (see Fig. \ref{fig.12}).
The broadening can be explained by elastic scattering of the produced protons by the abundant
$\rm{H}_2$ molecules in the source. Furthermore the yield of the proton source is three orders of magnitude higher than at the lower pressure.

\subsection{Electrostatic focusing and pulsed electric deflection}
\label{sec-4}

Various focusing elements are needed for the beam transport of the \rm{H}$^{-}$ ions leaving the Ar cell in the neutron BoB experiment. Einzel lenses are the best choice inside the throughgoing beam tube, whereas outside the biological shield an electrostatic quadrupole doublet can be used.

\begin{figure}[h]
\includegraphics[width=0.49\textwidth]{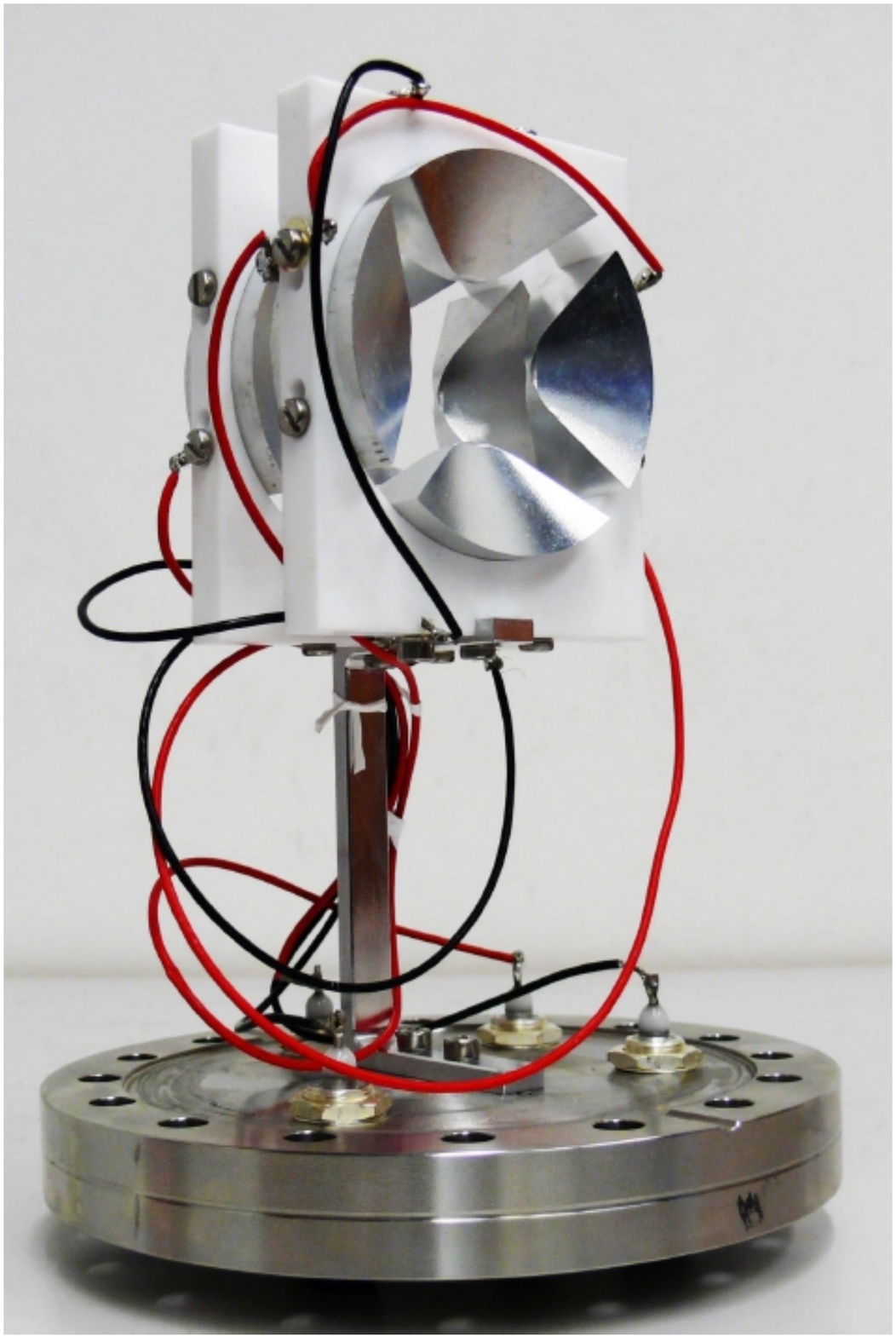}
\center
\caption{Electric quadrupole doublet.}
\label{fig.13}
\end{figure}

\begin{figure}[h]
\includegraphics[width=0.49\textwidth]{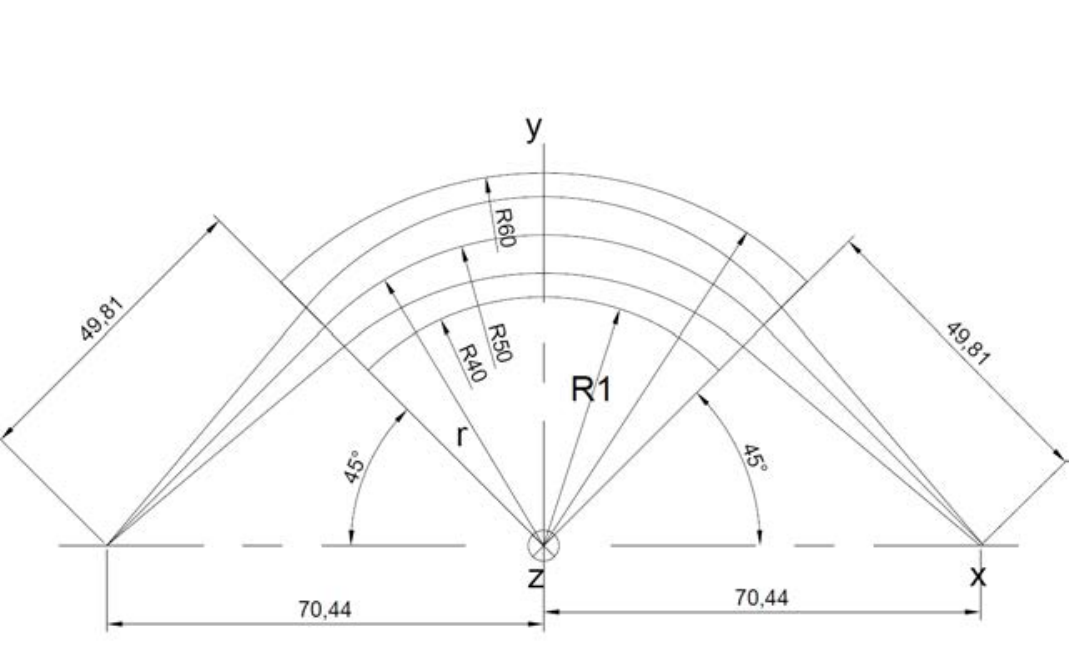}
\center
\caption{Sketch of a $90^{\ o}$ electric deflector consisting of two electrodes with curvature radii $R_1$ and $R_2$, $r$ being the reference particle radial coordinate. The deflector focuses both in horizontal and vertical directions.}
\label{fig.14}
\end{figure}

\begin{figure}[h]
\includegraphics[width=0.49\textwidth]{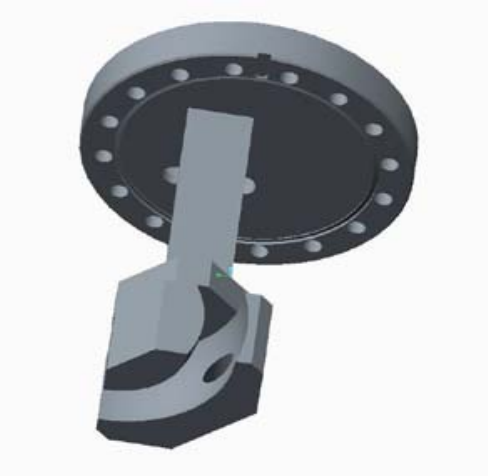}
\center
\caption{Pulsed electric deflector consisting of two spherically shaped insulated electrodes, one with a hole for the throughgoing beam.}
\label{fig.3}
\end{figure}

A photograph of a suitable quadrupole doublet (QPD) is shown in Fig. \ref{fig.13}. The device consists of two individual electric quadrupoles (QPs), which are operated in a crossed mode, resulting in a common focal point for all particle trajectories \cite{QPD}.
The focusing of the QPD is independent on the mass of the focused charged particle. Only the charge of the particle and the kinetic energy matters.
The QPD is compact and light, compared to magnetic systems. Furthermore it is cheap, concerning production costs.
The common focal point is achieved by operating the second QP at higher voltage than the first one \cite{QPD}.
This can be explained by a "simple" physical picture. For one focal area (A1), the first and the second QP act as a combination of a convex (first) and diverging lens (second). The focal length of this configuration is larger, compared to the focal length of the first QP for this area. For the second (crossed) focal area (A2) it is just the opposite case. The first QP acts as a diverging lens, while the second QP is a convex lens for this focal area (A2). The focal length for this focal area is also larger compared to pure focal length of the second QP. If both QP voltages are equal, then there is no common focal point possible \cite{Stef}. If the voltage of second QP is increased, compared to the first QP, then the total focal length of focal area A1 will increase (diverging strength of QP2 increases!). The focal length of A2 will decrease, because the focusing effect of QP2 will become stronger, due to the the higher voltage. Both focal areas will concur at a certain point \rm{z}$_f$, for a certain voltage of QP2, if the voltage of QP1 is fixed. The choice of voltages depends on the energy of the charged particle, and on the selected focal point \rm{z}$_f$.
We used this QPD in our BN gate test measurements (see Fig. \ref{fig.8}). There we placed the doublet after the BN gates, and focused the protons at the point, where the MCP detector was installed.
A further beam optic device, a deflector, which bends the particle trace of the \rm{H}$^-$ ions by $90^{\ o}$ was designed and built \cite{Hint} (see Fig. \ref{fig.3}).
Due to the spherical shaped electrodes, the radial and axial deflector focusing lengths are equal (see Fig. \ref{fig.14}).
The electric field of the deflector is

\begin{equation}
E(r)=\frac{U R_1 R_2}{r^2 (R_2-R_1)}.
\label{eq.7}
\end{equation}
The bending electric field $E$ for 500 eV protons at $r=\ 5\ \rm{cm}$ is $E=2\cdot\ 10^4\ \rm{V/m}$. The corresponding voltage is $U=416.7\ \rm{V}$, which means $+\ 208.4\ \rm{V}$ at the outer, and $-\ 208.4\ \rm{V}$ at the inner electrode.

\begin{figure}[h]
\includegraphics[width=0.49\textwidth]{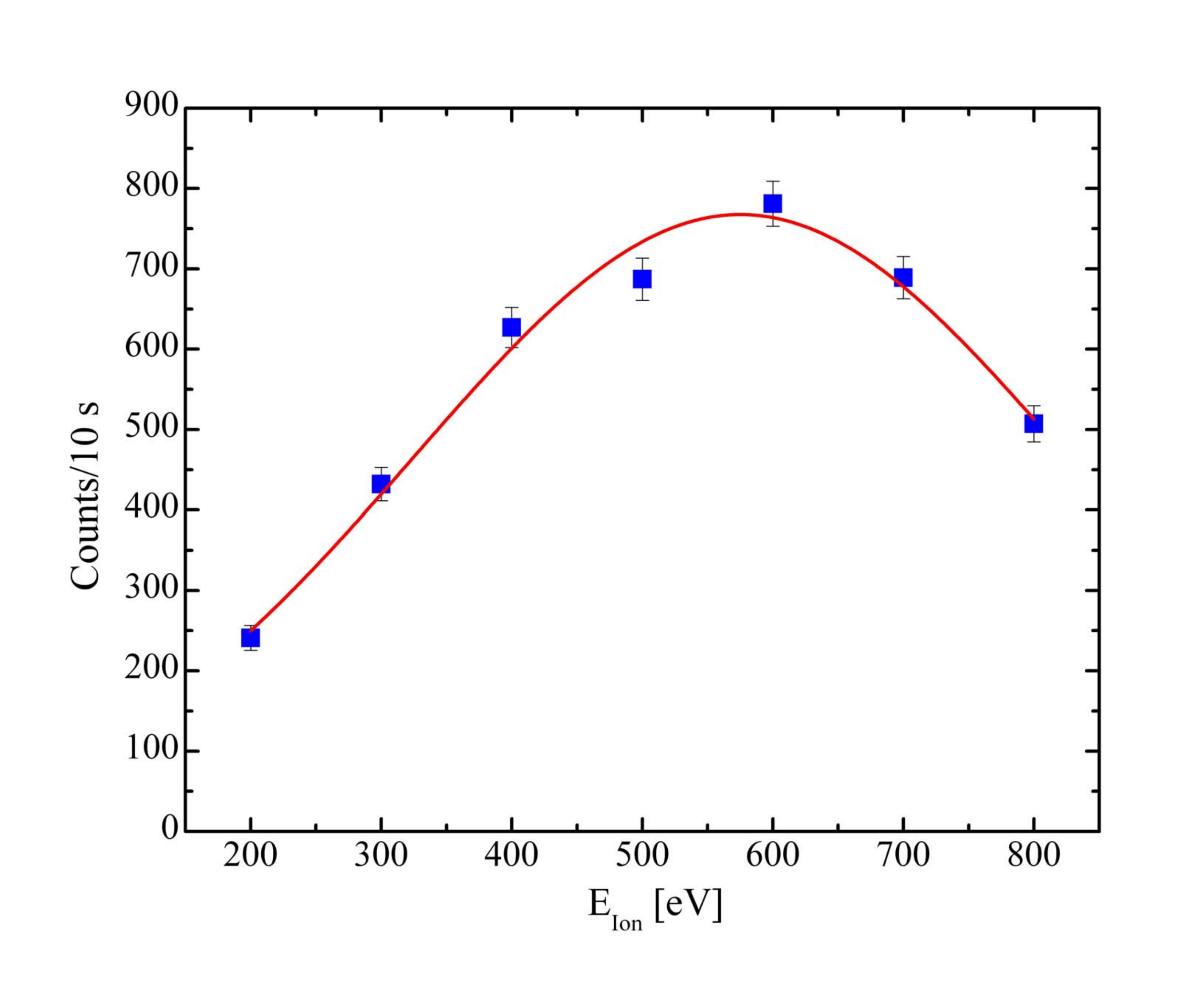}
\center
\caption{Dispersion of the electric deflector for O$^+$ ions.}
\label{fig.15}
\end{figure}

The deflector was successfully tested at our beam facility with O$^+$ ions \cite{Zusanna}. The results of these measurements (see Fig. \ref{fig.15}) also hold for protons,
because the electric deflection depends only on the charge and the kinetic energy of the ion. The dispersion of the deflector was measured by changing the kinetic energy of the  \rm{O}$^+$ ions.

\begin{figure}[h]
\includegraphics[width=0.49\textwidth]{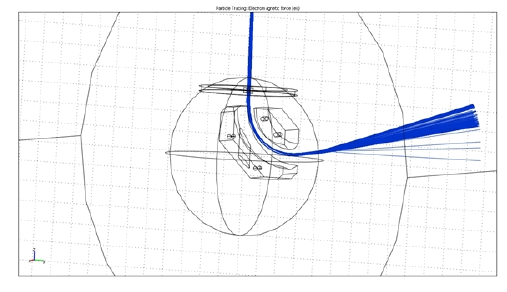}
\caption{Simulation of 500 eV proton (blue lines) through the electric deflector ($U_{\pm\ }=208\ \rm{V}$). The divergence of the beam is 1.4$\%$ (velocity).}
\label{fig.16}
\end{figure}

The deflector voltage $U_\pm$ was set to $\pm\ 208\  \rm{V}$, which deflects ions with 500 \rm{eV} kinetic energy. The data were fitted with a Gaussian function. The dispersion peaks at $E_0=575\pm8\ \rm{eV}$. The FWHM is $500\pm20\ \rm{eV}$\ ($\chi^2/DOF=1.4;\ R^2=0.98$).
The wide dispersion is remedied using the BN gate TOF system (or counter-field system) for the selection of the BoB H$^-$ ions.
COMSOL Multiphysics$^{tm}$ \cite{Comsol} simulations (see Fig. \ref{fig.16}) confirmed approximately the theoretical model \cite{Hint} leading to Eq. \ref{eq.7}.

\subsection{BN gate TOF chopper applications}
\label{sec-4}
A proton detection system, which uses secondary electrons from thin foils, produced by protons with typical energies in the region 5- 20 keV, was investigated with our BN gate TOF system \cite{Karina}.
This proton detector shall be used in experiments studying the decay of the free neutron, were protons occur in the energy range of 0 - $\sim$750 eV \cite{Nico}.
These protons are accelerated to high higher energies by applying high voltage, and converted to secondary electrons, which can be detected by standard electron detectors.

\begin{figure}[h]
\includegraphics[width=0.49\textwidth]{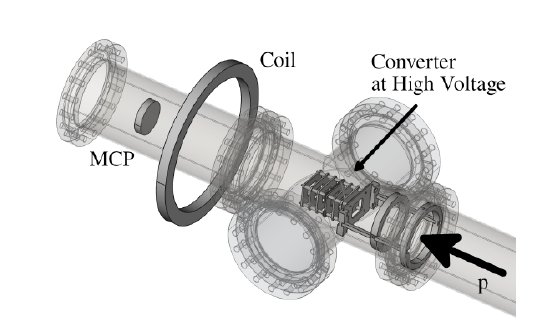}
\caption{Experimental setup for measuring the secondary electron production in thin foils by protons.
Different foils were tested (carbon foils coated with MgO and LiF). The foils are placed in the center of the degrader.}
\label{fig.17}
\end{figure}

\begin{figure}[h]
\includegraphics[width=0.49\textwidth]{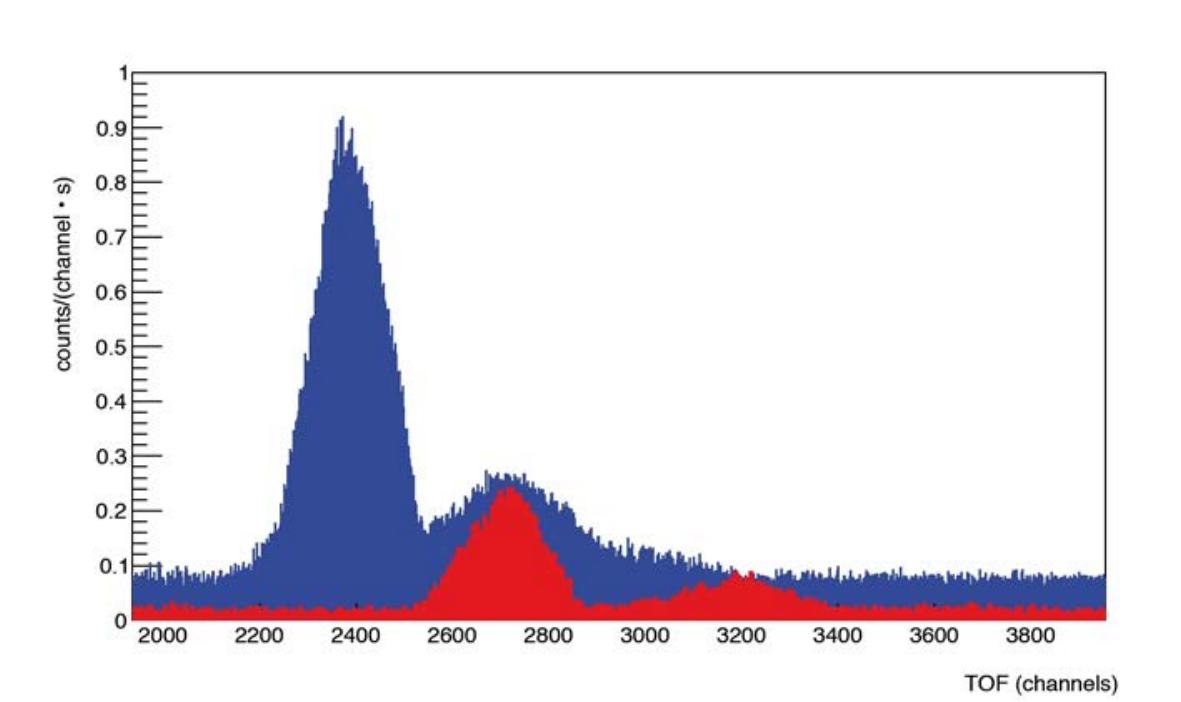}
\caption{Blue: Secondary electron production by 500 eV protons, accelerated to 18 keV at the foil position in the degrader.
Red: Proton  (500 eV) distribution at 0 V voltage in the degrader. The proton source H$_2$ pressure was 3$\cdot$10$^{-3}$ mbar.}
\label{fig.19}
\end{figure}

The measurements were performed at our proton source in the BoB lab.
The degrader system ( see Fig. \ref{fig.17}) was installed 0.58 m (entrance frame) after the second BN gate (see Fig. \ref{fig.10}). The distance of the exit frame to the MCP detector was 0.28 m.
As an example Fig. \ref{fig.19} shows a measurement with a carbon foil (17 $\mu$g/cm$^2$), coated with 10 $\AA$ LiF.

The peak position (TOF channel 2370 in Fig. \ref{fig.19}) of the secondary electrons correspond to a kinetic energy of 18.5 keV of the secondary electrons.
A simple approximative calculation shows, that $\sim$29$\%$ of the secondary electrons reach the MCP detector. The detection efficiency of 18 keV electrons is around 20$\%$.
The proton detection efficiency for 500 eV protons is 5$\%$ \cite{Hama}.
Using these rough estimates, we get a gain of 3.1 secondary electrons per incident proton.
Thorough studies of the electric field distribution and further measurements including incident beam intensity monitoring are planned to obtain more accurate secondary electron yield values.

\section{Conclusion}
\label{sec-5}
A new experiment to detect the bound beta neutron decay $n\rightarrow\textrm{H}\overline{\nu}_e$ requires the development of
novel methods and technologies. The key requirement is a high rejection of broad band backgrounds from thermal and fast hydrogen atoms of yet unknown magnitude. It results from both the rest gas and hydrogen forming from the abundant neutron decay protons picking up an electron. We have demonstrated these technologies using a laboratory setup producing fast protons and hydrogen atoms at energies up to 500 eV. A system of Bradbury Nielsen gates was built and used as TOF selector system (chopper) for a narrow band beam of fast protons. Such a fast switching electric system can also be used for other measurements in neutron decay. We have  operated an Argon filled gas cell to verify the charge exchange process. This cell still needs to be optimized concerning the number density of Argon atoms. We have also simulated, built and tested the dispersive ion optics and the ion detector using protons and positive Oxygen ions. This system is to be placed downstream of the Argon cell acting as a second velocity discriminating system. With these technologies at hand, we can now optimize the full set-up to be operated with metastable hydrogen atoms in order to probe shielding of electric fields and to verify the efficiencies of individual components. We thus are conceptually prepared to build a first experiment at a high flux neutron source with throughgoing beam pipe.

\section{Acknowledgment}
\label{sec-6}
Many thanks to Daryl Bishop, TRIUMF, for the pulse generating electronics design support and to S. Winkler (TUM) and P. Hartung (LMU) for valuable advice and assistance.
This work was supported by the cluster of excellence "Universe"  and by the
Maier-Leibnitz-Laboratorium (MLL) of the
Ludwig-Maximilians-Universit\"at (LMU) and the Technische
Universit\"at M\"unchen (TUM).

%
%
%

\end{document}